\documentstyle[aas2pp4]{article}

\begin{document}

\title{Discovery of Kilohertz QPOs in the Atoll X-Ray Binary 4U~1705-44}

\authoremail{ecford@astro.uva.nl}

\author{Eric C. Ford\altaffilmark{1}, Michiel van der Klis\altaffilmark{1}, 
and Philip Kaaret\altaffilmark{2} }

\altaffiltext{1}{Astronomical Institute, ``Anton Pannekoek'',
University of Amsterdam, Kruislaan 403, 1098 SJ Amsterdam,
The Netherlands}
\altaffiltext{2}{Columbia Astrophysics Laboratory, 
Columbia University, 550 W. 120th Street, New York, NY 10027}

\begin{abstract}

In observations with the Rossi X-ray Timing Explorer we have
discovered quasi-periodic oscillations (QPOs) near 1 kHz from
4U~1705-44, a low--mass X-ray binary with a neutron star classified as
an atoll source. In six separate observations, we detect one QPO with
a frequency ranging between 770 and 870 Hz and a 4\% rms fraction in
the full detector energy band.  There is evidence for a second QPO at
1073 Hz in one interval. The separation in frequency of the two QPOs
is $298\pm11$ Hz.  The QPOs are present only in observations where the
mass accretion rate is inferred to be at an intermediate level, based
on the atoll source phenomenology.  At the highest accretion rates,
the QPOs are not detected with upper limits to the rms fraction of
about 2\%. At the lowest accretion rates the upper limits are about
4\%.  The QPO frequency increases with inferred mass accretion
rate. This is expected in models where the QPO frequency is generated
by motion at an inner edge of the accretion disk. An increased mass
accretion rate causes the disk edge to move in, increasing the orbital
frequency.  Five Type--I X-ray bursts are observed with no detectable
oscillations.

\end{abstract}


\section{Introduction}

Quasi-periodic oscillations (QPOs) with frequencies near 1 kilohertz
were discovered from X-ray binaries with the Rossi X-Ray Timing
Explorer (RXTE) almost immediately after its launch (for reviews and
references see van der Klis 1997, Kaaret \& Ford 1997a, Ford
1997). These QPOs are likely produced very near the accreting neutron
star where the dynamical time scale is near $10^{-3}$
seconds. Observations of such fast oscillations provides a new
opportunity to measure fundamental properties of neutron stars and the
effects of strong field gravity in low mass X-ray binaries.

Kilohertz QPOs have been discovered in both major subclasses of LMXBs:
the Z-sources and the atoll sources (see Hasinger \& van der Klis
1989). Certain trends are already apparent.  For example, the
kilohertz QPOs in Z-sources are much weaker than those in the lower
luminosity atoll sources. In the sources with luminosities
intermediate between the majority of Z and atoll sources, the
persistently bright galactic bulge sources, QPOs are apparently absent
or very weak (Wijnands, van der Klis, \& van Paradijs 1997a). To
understand the physical mechanisms at work, further observations of a
large sample of sources are needed with the unique capabilities of
RXTE, likely to be unequaled in the next decade.

Here we report the discovery of fast QPOs from the low mass X-ray
binary 4U~1705-44 using RXTE.  4U~1705-44 has been classified as an
atoll source (Hasinger \& van der Klis 1989). Its different source
states are supposed to reflect changes in mass accretion rate. We
observe a link between the source state and the presence and
frequencies of fast QPOs.

In Section~2 we summarize the observations and analysis techniques.
Section~3 presents the detections of fast QPOs and X-ray burst
properties.  In Section~4 we identify the atoll source states during
our observations. Section~5 is a discussion of the relation of the
source states to the QPO properties and physical implications.

\section{Observations \& Analysis}

We have observed 4U~1705-44 with RXTE at random times from February to
June 1997. The total usable observing time amounts to 55.3 ksec,
broken into 18 separate continuous intervals.  Here we use data from
the Proportional Counter Array (PCA), (Zhang et al. 1993), which has
high time resolution good sensitivity above background from about 2 to
30 keV.  In these observations all five detector units of the PCA are
on.

The All-Sky Monitor on RXTE shows that during our observations the
flux of 4U~1705-44 is smoothly varying and has a minimum at about 1
April 1997.  In our observations the count rate in the PCA varies from
500 to 2800 c/s, with a flux of $0.7-5.3\times10^{-9}$
erg~cm$^{-2}$~s$^{-1}$ (2--10 keV).

In all observations we have initiated an `Event' mode which provides
data in time intervals of 122 $\mu$sec in 64 energy channels ranging
from about 1 to 100 keV.  From this data, we extract average fourier
power density spectra in 64 second data intervals. The minimum and
maximum observable frequencies are thus 0.016 and 4096 Hz.

Since QPO strengths depend on energy, it is important to consider
energy subbands in searching for QPOs.  We employ three energy bands:
2.4--6.4 keV, 6.4--31 keV, and the full energy range of the PCA.  We
fit each QPO feature in the power spectra with a Lorentzian plus a
constant. The fit values determine the centroid frequencies, widths
and rms fractions of the QPOs. We use $\Delta\chi^{2}=1$ for quoting
errors and $\Delta \chi^{2}=2.71$ for upper limits (95\% confidence).
Only those QPO features with significance greater than $3\sigma$ are
reported here.

\section{Fast QPO Detections}

\begin{figure*}
\figurenum{1}
\epsscale{1.8}
\plotone{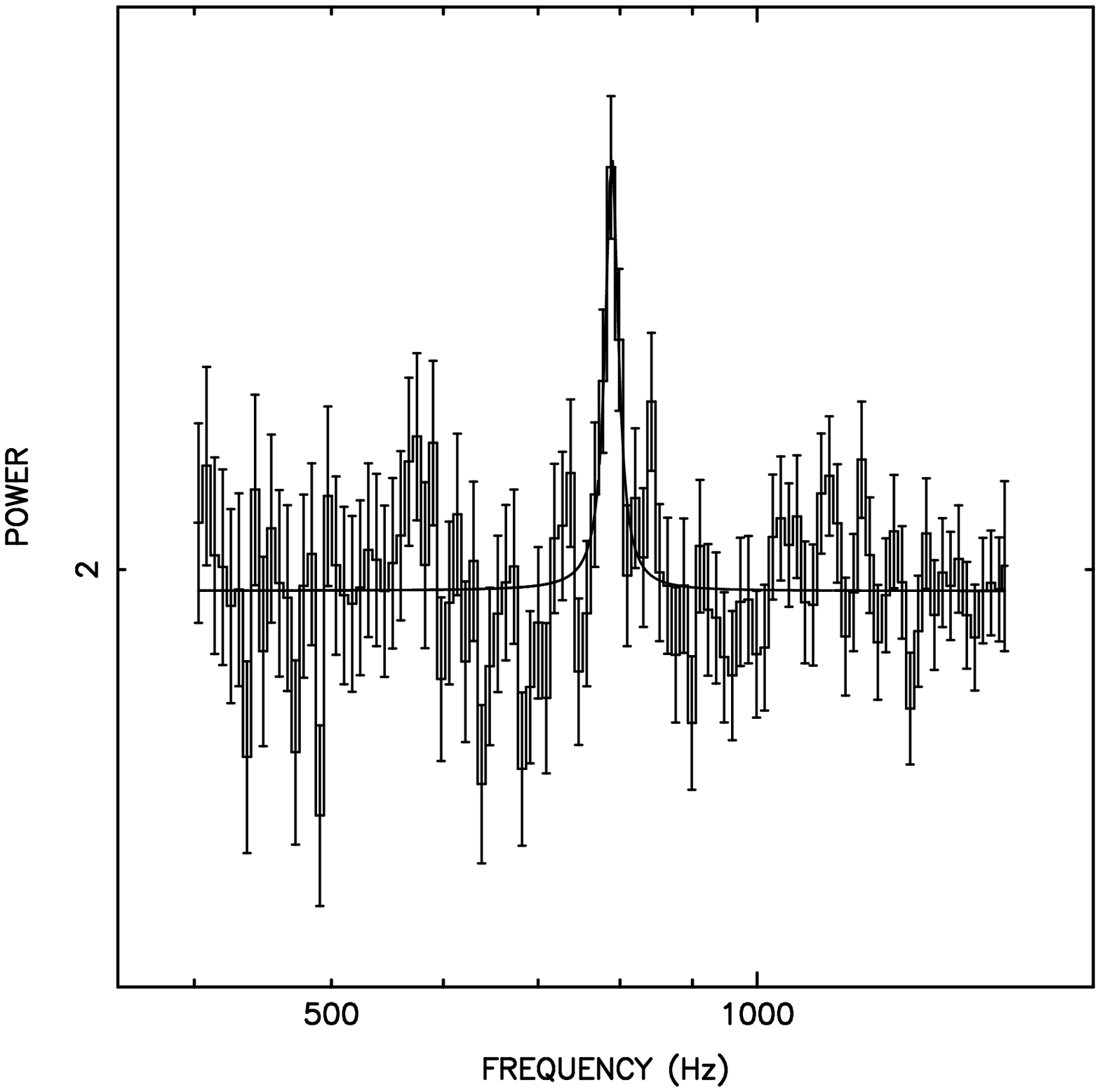} 
\caption{Power density spectrum from the 3072 sec observing interval 
beginning 1997 April 3 18:23:56 UTC in the energy range 6.4--31 keV. The fit 
is a Lorentzian with a centroid frequency of $792\pm3$ Hz, and rms 
fraction $4.4\pm0.5$\% plus a constant.}
\label{fig:pds}
\end{figure*}

\begin{figure*}
\figurenum{2}
\epsscale{1.8}
\plotone{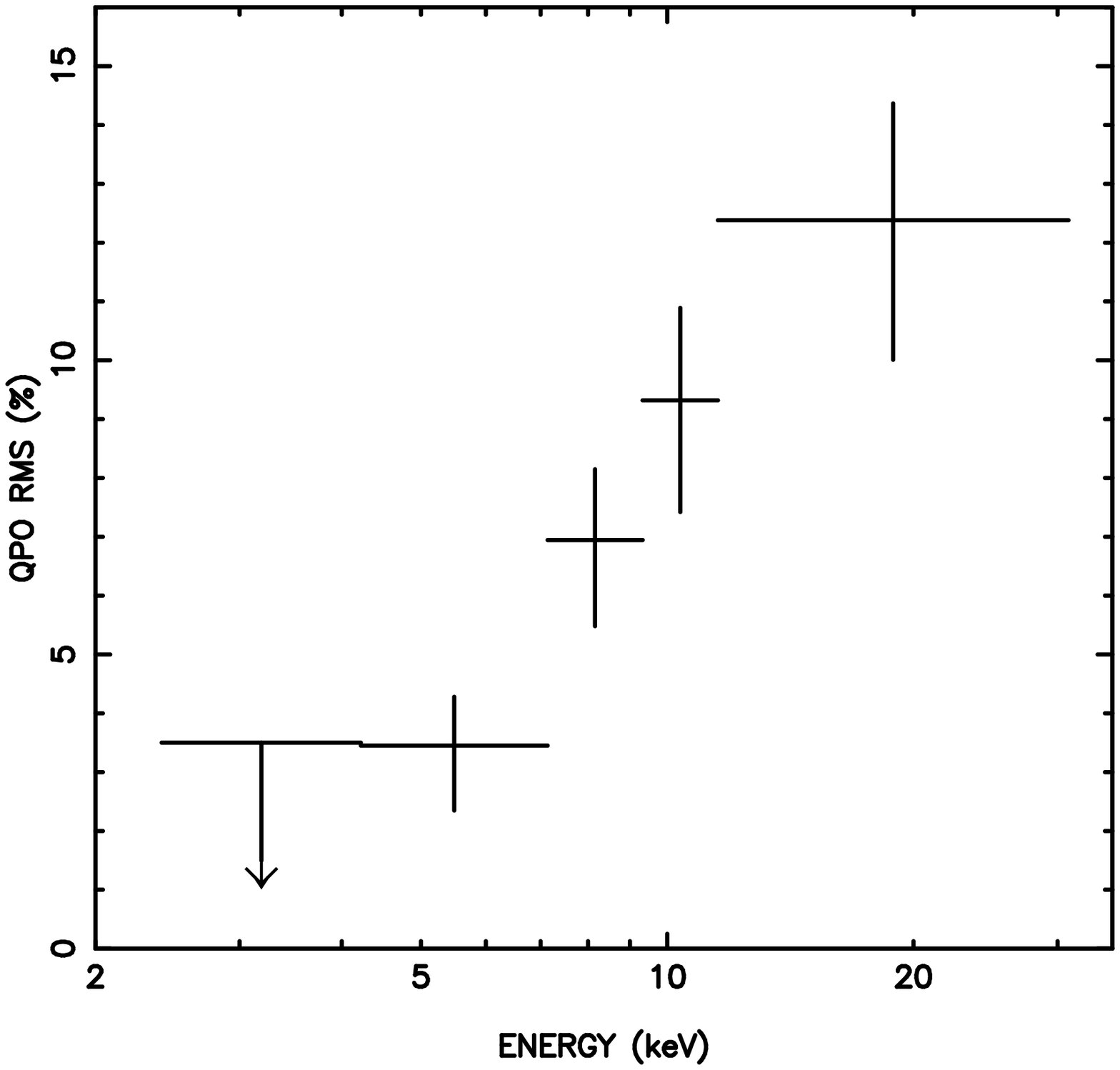}
\caption{Fractional rms vs. energy of the 792 Hz QPO shown in
Figure~1.  Error bars use $\Delta \chi^{2}=1$. The upper limit is 95\%
confidence.}
\label{fig:rmsE}
\end{figure*}

Figure~\ref{fig:pds} shows an example power spectrum from 3072 seconds
of data beginning 1997 April 3 18:23:56 UTC. The narrow QPO at
$790\pm3$ Hz is detected with a significance of $5.4\sigma$ for a
single fitting. Six of our 18 observing intervals exhibit QPOs in the
range 780 to 870 Hz (Table~1).  Most QPOs are detected in both the
6.4--31 keV band and the full energy band, but not in the 2.4--6.4 keV
band. The rms fraction of the QPOs increase with energy
(Figure~\ref{fig:rmsE}), similar to other sources.

In addition to the QPO clearly detected at 780 to 870 Hz, we have
evidence for a second QPO at higher frequency. This QPO is visible in
one observing interval (1997 April 3 16:38:04) at $1074\pm10$ Hz in
the 2.4--6.4 keV band with an rms fraction $3.1\pm0.5$\%. The
significance is $5.2\sigma$, though this is for a single trial. The
quoted error in frequency includes systematic errors due to rebinning
the power spectra in frequency, an important effect for weak features.
The two QPOs in this interval are separated by $298\pm11$~Hz.

To help confirm the presence of the second QPO we have implemented the
`shift--and--add' technique of Mendez et al. (1998) with which one can
improve sensitivity by adding multiple power spectra.  We take all the
power spectra where the lower frequency QPO is detectable, rescale the
frequencies so the lower frequency QPOs all have the same centroid,
and then add the power spectra together.  With this technique, we also
detect the second QPO.  The significances and rms fractions are:
$3.6\sigma$, $3.1\pm0.5$\% (full energy band); $3.2\sigma$,
$2.6\pm0.5$\% (2.4--6.4 keV); and $2.9\sigma$, $3.1\pm1.7$\% (6.4--31
keV).  The upper limits for a typical single observing interval are
3.1\% (full band) and 3.9\% (6.4--31 keV). Thus, our non-detection in
single intervals is consistent with the rms fractions determined when
the data are combined using the shift technique.  In all energy bands
the frequency separation of the two QPOs is consistent with with the
previously quoted value of $298\pm11$ Hz to within $1.6\sigma$.

There are five X-ray bursts during our observations.  We have searched
for QPOs in all of the bursts, generating power spectra in one--second
intervals in each of the three energy bands identified above. We
detect no QPOs in any of the time intervals or energy subbands. The
upper limits to the rms fraction at the peaks of the bursts are
approximately 5\% in the three weak bursts and 2\% in the stronger
bursts.  These upper limits are much smaller than the amplitudes of
the QPOs in the bursts from 4U~1728-34 where the amplitudes reach 20\%
and perhaps even 40\% (Strohmayer, Zhang \& Swank 1997).

Of the five bursts from 4U~1705-44, two occur when the source is in
the banana state (16 February and 4 April). These bursts have total
durations of about 10 sec, rise times of 0.6 sec, and average
temperatures of about 1.1~keV. Three additional bursts occur when the
source is in the island state (1 April). These are about 3 times
longer, have peak count rates about 4 times smaller, rise times of
4.5--5.0 sec, and average temperatures of about 1.3~keV.

The bursts in 4U~1705-44 are cooler compared to those in 4U~1728-34
(roughly 1.2 vs. 2.2 keV) and have smaller fluences and rise times.
The absence of QPOs in 4U~1705-44 bursts is perhaps due to the lower
temperatures, and larger inferred sizes (6--12 km radii for a distance
of 11 kpc). If the oscillation is due to burning in a hot spot
(Strohmayer, Zhang \& Swank 1997) a larger emitting area results in a
reduced oscillation strength. Oscillations may also be absent if the
hot spot is near the rotation axis or that our line of sight is
closely aligned with the spin axis of the neutron star.

\section{Source States}

\begin{figure*}
\figurenum{3}
\epsscale{1.8}
\plotone{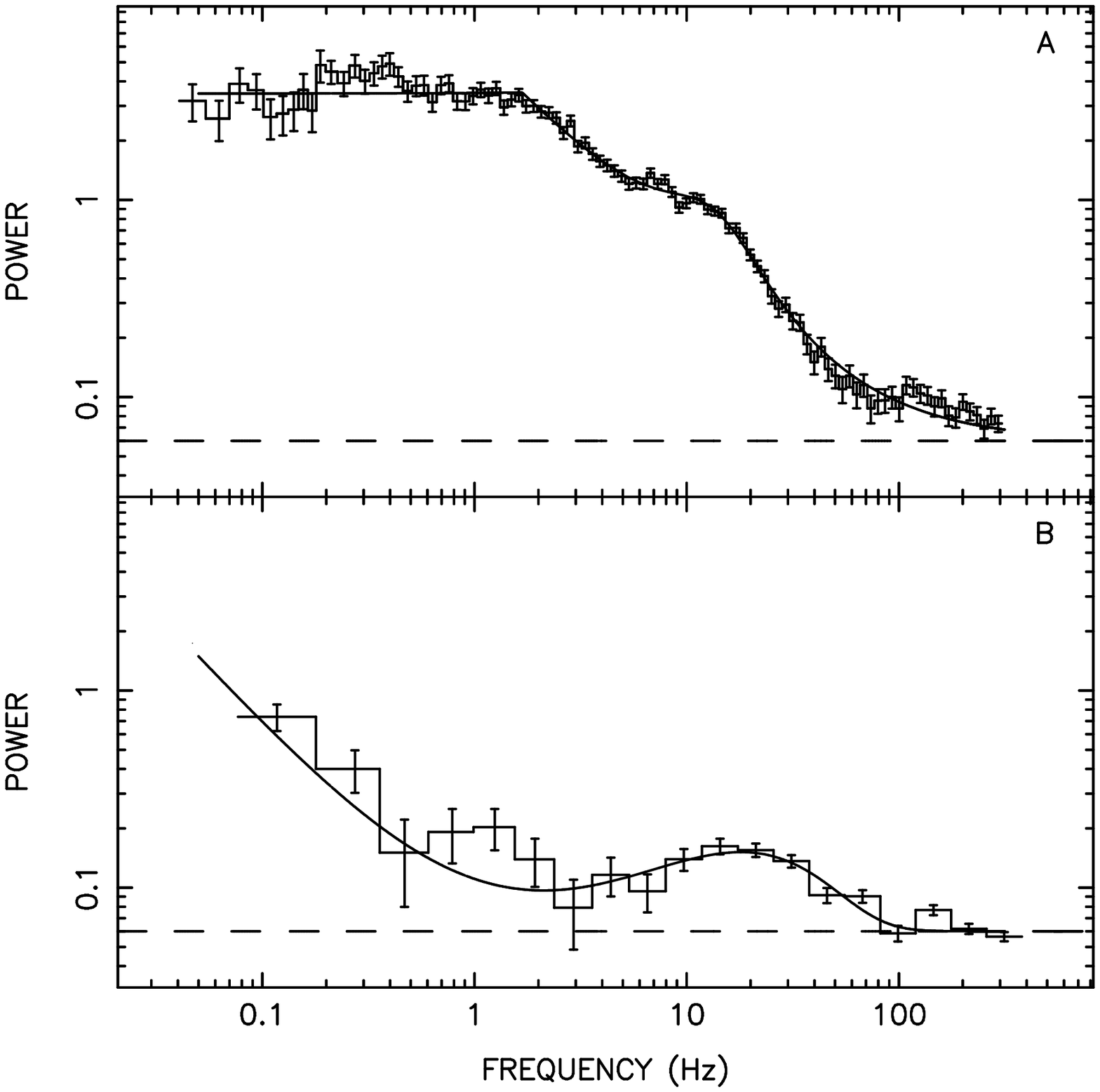} 
\caption{Low frequency power density spectra for the entire PCA energy
band in the intervals beginning {\em a}) 1997 April 1 15:01:25 and
{\em b}) 1997 April 4 8:37:13 UTC.  The spectra are Leahy
normalized. A constant has been subtracted and the Poisson levels are
shown by the dashed lines.  Observation {\em a} is an `Island' state
with a total rms fraction of the excess noise of 21.4\%.  (0.1--100
Hz). Observation {\em b} is a `Banana' state with an rms fraction of
6.1\%.  The fits are standard functions for atoll sources (Hasinger \&
van der Klis 1989).  In {\em a}, a broad Lorentzian is added with a
centroid at 10 Hz.}
\label{fig:lowf}
\end{figure*}

For each observation, we have identified the source state following
Hasinger \& van der Klis (1989), based on the power at low frequencies
(approximately 0.01 to 100 Hz) and the X-ray colors of the source. At
low count rates (29 March and 1 April), the source is in an `island'
state of the atoll source; below about 10 Hz there is strong
band--limited noise, with a fractional rms greater than 20\%
(Fig~\ref{fig:lowf}{\em a}).  As the count rate increases, the source
moves into a `banana' state, the band--limited noise gets much weaker,
and a separate noise component appears in the power spectra below 1 Hz
described by a power law (Fig~\ref{fig:lowf}{\em b}). The observations
on 16 February, 3 April and 4 April are in the banana state. On 9 June
the source is in the upper banana.

There are three days of observations in the lower banana state: 16
February, 3 April and 4 April. Within the lower banana, the hard color
of the source gets softer while the high-frequency noise component
decreases. This behavior is consistent with the motion along the
banana branch as observed in 4U~1705-44 with EXOSAT (Hasinger \& van
der Klis 1989), though in this lowest part of the banana branch the
movements in the color diagrams are very small and the changes in the
amplitude of the noise components are significant only at the
$1\sigma$ level. In order of increasing progress along the banana
branch, the observations are ordered as: 3 April, last half of 4
April, first half of 4 April, and 16 February.  In Table~1 we assign a
numerical ranking of the states from 1 to 7 going from the island
state to the upper banana state. The inferred mass accretion increases
from 1 to 7.

\section{Discussion}

The 760--870 Hz QPO in 4U~1705-44 is probably a single feature and is
likely the lower frequency of two QPO peaks. The second QPO, which we
identify at 1074 Hz, then is the higher frequency peak. More
observations in the lower banana state are needed to confirm the
presence of this second QPO. Such phenomenology is very similar to
other atoll sources. A single strong QPO near 800 Hz is present at
times in several sources, e.g. 4U~1608-52 (Berger et al. 1996; Mendez
et al. 1998), 4U~1820-30 (Smale, Zhang \& White 1997), 4U~1636-53
(Zhang et al. 1996; Wijnands et al. 1997b). In all of these sources a
second QPO is sometimes detectable (see above references). The
frequency separation of the two peaks is similar to that in
4U~1705-44.  Such double QPOs, and also the QPOs observed in X-ray
bursts (e.g. Strohmayer, Zhang \& Swank 1997), suggest a beat
frequency mechanism (Alpar \& Shaham 1985; Miller, Lamb \& Psaltis
1998). If such an interpretations holds, then the frequency separation
of the QPOs in 4U~1705-44 implies that the spin period of the neutron
star in this system is $3.35\pm0.12$ msec.

The appearance and frequency of the fast QPOs in 4U~1705-44 are
correlated with the states of the source. The states in turn are
thought to be related to the mass accretion rate, with the accretion
rate increasing from the island to the banana.  At the lower end of
the banana branch a QPO appears at about 780 Hz.  Further along the
branch the frequency increases to 835 Hz and then 865 Hz.  This
behavior supports models where the QPO modulation is generated at an
inner disk edge which shrinks as the mass accretion rate increases
(e.g. Miller, Lamb \& Psaltis 1998).  We detect the second QPO at 1073
Hz only in the the lowest part of the banana where the frequency of
the stronger QPO is at its lowest.  Farthest along the banana branch
the QPO becomes undetectable with rms fraction upper limits of about
2\%.  At the highest accretion rates, the reduced QPO amplitude is
perhaps due to spreading of the accretion stream, alternatively a
puffed--up inner disk may obscure the central source (Smale, Zhang \&
White 1997).

Similar behavior is observed in other X-ray binaries.  Atoll source
state identifications have been made simultaneous with fast QPO
detection in 4U~1636-53 (Zhang et al. 1996; Wijnands et al.  1997b),
4U~1735-44 (Wijnands et al. 1997c), 4U~1820-30 (Smale, Zhang \& White
1997) and KS~1731-260 (Wijnands \& van der Klis 1997d).  These sources
were observed in the banana branch. Consistent with the present data,
the QPO in all cases is detected only in the lower part of the banana
and disappears as the source moves up along the banana branch. In
4U~0614+091 a similar effect appears as the QPO amplitude decreases as
the count rate increases (Ford 1997).

The situation is less clear at low mass accretion rates, i.e. in the
island states.  Our upper limits of about 4\% rms fraction are not
very constraining.  QPOs have been detected in island states in both
4U~0614+091 (Mendez et al. 1997) and 4U~1608-52 (Yu et al. 1997). At
the lowest count rates in the island state of 4U~0614+091 the QPO is
not detected (Mendez et al. 1997). This fits the general trend we
observe here.

The QPO frequency shows no correlation with count rate or energy flux
in the 2--10 keV band, though we note that the range of count rates is
small (1205 to 1236 c/s). An effect similar to that seen in
4U~0614+091 may be at work, where there is no unique correlation
between rate and frequency in different observations separated by
several months (Ford et al. 1997a). The present data indicate that the
QPO frequency is better correlated with the atoll state of the
source. Changes in the source state are reflected as changes in the
x-ray colors, which in turn are simply changes in the energy spectrum
of the source. Therefore the state--frequency correlation also
manifests as a spectrum--frequency correlation.  From other sources we
know that the changes in QPO frequency are linked to changes in the
energy spectrum.  The frequencies of the QPOs in 4U~0614+091 are well
correlated with the flux of a blackbody component in the energy
spectrum (Ford et al. 1997b), and in that source and also 4U~1608-52
the QPO frequencies are correlated with the spectral index of the
power law component in the energy spectrum (Kaaret et al. 1998).

We thank Mariano Mendez for assistance in implementing the frequency
shifting technique.  We thank Rudy Wijnands, Rob Fender and the
referee for helpful discussions and comments. ECF acknowledges support
by the Netherlands Foundation for Research in Astronomy with financial
aid from the Netherlands Organization for Scientific Research (NWO)
under contract numbers 782-376-011 and 781-76-017.


\begin{figure*}
\figurenum{}
\epsscale{2.1}
\plotone{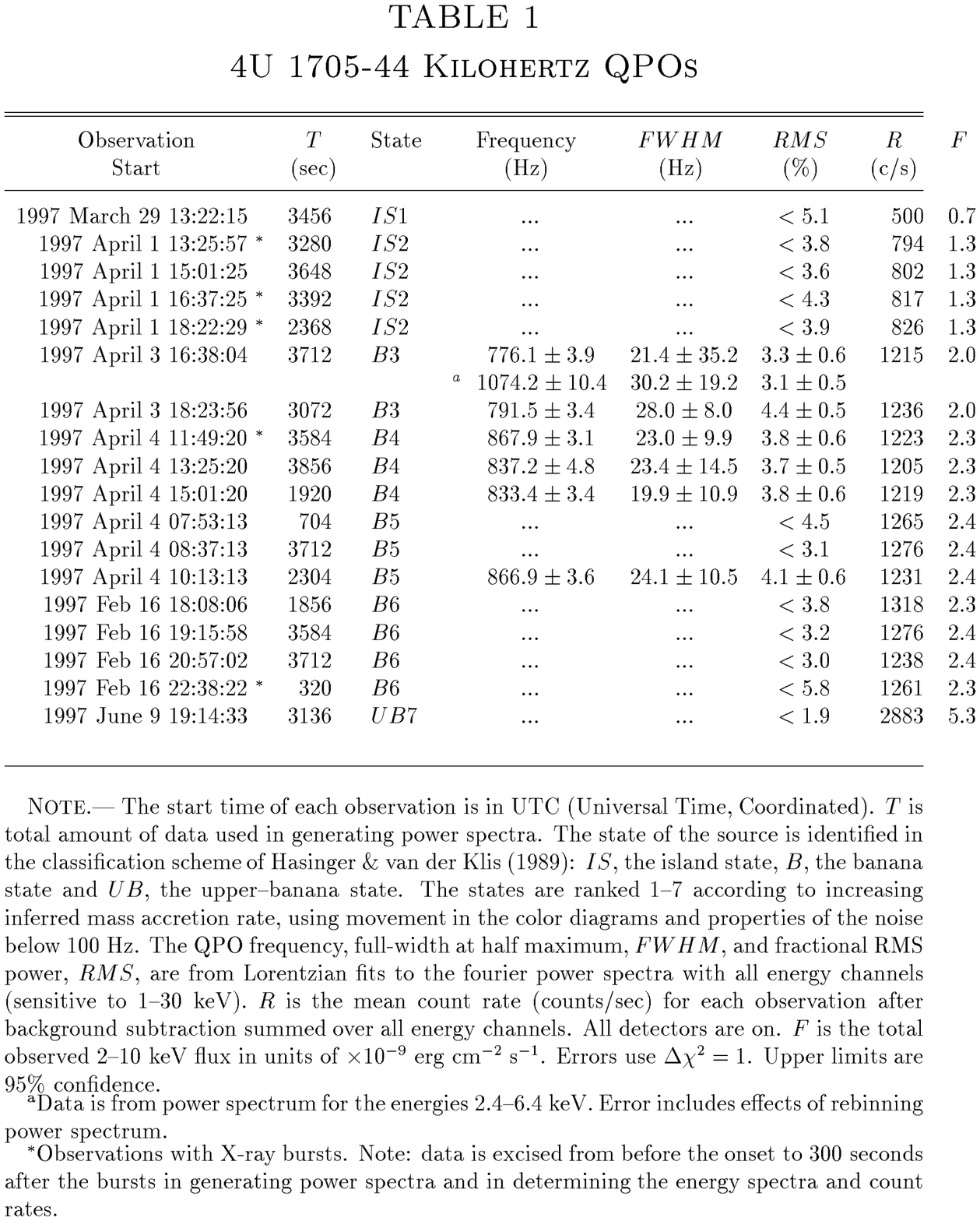} 
\caption{ }
\label{tbl:obs}
\end{figure*}



\begin{references}

\reference{alpar85} Alpar, M.A., \& Shaham, J., Nature, 316,
239-241 (1985)
\reference{berger96} Berger, M., et al., ApJL, 469, L13-16 (1996)  
\reference{christian97} Christian, D.J. \& Swank, J.H., ApJS, 109
177-224 (1997)
\reference{ford97a} Ford, E.C. et al., ApJL, 475, L123-126 (1997a)
\reference{ford97a} Ford, E.C. et al., ApJL, 486, L47-50 (1997b)
\reference{ford97b} Ford, E.C., Ph.D. Thesis, Columbia University (1997)
\reference{hasinger89} Hasinger, G. \& van der Klis, M., AA, 225, 79-96 (1989)
\reference{kaaret97a} Kaaret, P., \& Ford, E.C., Science, 
276, 1386-1391 (1997a) 
\reference{kaaret98} Kaaret, P. et al., submitted to ApJL (1998)
\reference{mendez98} Mendez, M., et al., ApJL, 494, L65-69 (1998)
\reference{mendez97b} Mendez, M., et al., ApJL, 485, L37-40 (1997)
\reference{miller97} Miller, M.C., Lamb, F., and Psaltis, D., 
astro-ph/9609157, ApJ in press (1998)
\reference{smale97} Smale, A.P., Zhang, W. and White, N.E., ApJL, 483, 
L119-122 (1997) 
\reference{strohmayer97} Strohmayer, T., Zhang, W., and Swank, J., ApJL, 
487, L77-80, (1997)
\reference{vdk97} van der Klis, M., astro-ph/9710016, to appear in 
Proc. NATO Advanced Study Institute ``The many faces of neutron stars", 
Lipari, Italy (1997)
\reference{wijnands97a} Wijnands, R.A.D., van der Klis, M. and van 
Paradijs, J., astro-ph/9711222, to appear Proceedings of IAU Symposium 
188 ``The Hot Universe" (1997a)
\reference{wijnands97b} Wijnands, R.A.D. et al., ApJL, 479, L141-144 (1997b)
\reference{wijnands97c} Wijnands, R.A.D. et al., ApJL, 495, L39-42 (1998)
\reference{wijnands97d} Wijnands, R.A.D. and van der Klis, M. 
ApJL, 482, L65-68, (1997d)
\reference{yu97} Yu, W., et al., ApJL, 490, L153-165 (1997) 
\reference{zhang93} Zhang, W., et al. SPIE, 2006, 324-333 (1993)
\reference{zhang96} Zhang, W., Lapidus, I., White, N.E. \& 
Titarchuk, L., ApJL, 469, L17-19 (1996) 

\end{references}
\end{document}